\documentclass[11pt,a4paper]{article}
\pdfoutput=1
\usepackage{jheppub}
\usepackage{tikz}
\usepackage{subcaption}
\DeclareMathOperator{\Tr}{Tr}

\newcommand{\ri}{\mathrm{i}}

\newcommand{\hf}{\frac{1}{2}}
\newcommand{\qu}{\frac{1}{4}}
\newcommand{\til}[1]{\widetilde{#1}}

\newcommand{\del}{\partial}

\newcommand{\bra}{\langle}
\newcommand{\ket}{\rangle}

\newcommand{\bt}{\beta}

\newcommand{\rt}[1]{\sqrt{#1}}
\newcommand{\cO}{\mathcal{O}}

\newcommand{\cF}{\mathcal{F}}

\newcommand{\cD}{\mathcal{D}}

\newcommand{\cB}{\mathcal{B}}

\newcommand{\bbZ}{{\mathbb Z}}
\newcommand{\Fo}{F^{\rm o}}
\newcommand{\gs}{g_{\rm s}}
\newcommand{\dg}{{\tilde{g}}}
\newcommand{\zs}{z_*}
\newcommand{\xis}[1]{\xi_*^{(#1)}}
\newcommand{\nD}{D}
\newcommand{\tZ}{\widetilde{Z}}
\newcommand{\tW}{\widetilde{W}}
\newcommand{\cK}{\mathcal{K}}

\DeclareMathOperator{\Erfc}{Erfc}
\newcommand{\bket}[1]{\langle{#1}\rangle}

\begin{document}

\title{'t Hooft expansion of
multi-boundary correlators in 2D topological gravity}

\author[a]{Kazumi Okuyama}
\author[b]{and Kazuhiro Sakai}

\affiliation[a]{Department of Physics, Shinshu University,\\
3-1-1 Asahi, Matsumoto 390-8621, Japan}
\affiliation[b]{Institute of Physics, Meiji Gakuin University,\\
1518 Kamikurata-cho, Totsuka-ku, Yokohama 244-8539, Japan}

\emailAdd{kazumi@azusa.shinshu-u.ac.jp, kzhrsakai@gmail.com}

\abstract{
We study multi-boundary correlators
of Witten--Kontsevich topological gravity in two dimensions. 
We present a method of computing an open string like expansion,
which we call the 't Hooft expansion,
of the $n$-boundary correlator for any $n$
up to any order by directly solving the Korteweg--De Vries equation.
We first explain how to
compute the 't Hooft expansion of the one-boundary correlator.
The algorithm is very similar to that for the genus expansion
of the open free energy.
We next show that the 't Hooft expansion of correlators
with more than one boundary can be computed algebraically
from the correlators with a lower number of boundaries.
We explicitly compute the 't Hooft expansion of the $n$-boundary
correlators for $n=1,2,3$. Our results reproduce
previously obtained results for Jackiw--Teitelboim gravity
and also the 't Hooft expansion of
the exact result of the three-boundary correlator
which we calculate independently in the Airy case.
}

\maketitle

\section{Introduction}\label{sec:introduction}
In a recent paper \cite{Saad:2019lba} it was shown that the path integral of
Jackiw--Teitelboim (JT) gravity \cite{Jackiw:1984je,Teitelboim:1983ux}
is equivalent to a certain double-scaled random matrix model.
The genus expansion of this random matrix model describes the splitting/joining of the baby universes \cite{Saad:2019lba}.
In this correspondence we can consider 
the average $\bra Z(\bt)\ket$ of the partition function
$Z(\bt)=\Tr e^{-\bt H}$ where the average is defined by the integral over the random matrix
$H$.
More generally, we can consider the multi-point function
$\bra \prod_{i=1}^n Z(\bt_i)\ket$ of the partition functions $Z(\bt_i)~(i=1,\ldots,n)$.
On the bulk gravity side it corresponds to
the multi-boundary correlator, i.e.~the gravitational path integral 
on the spacetime with $n$ boundaries with fixed lengths $\bt_i$. 
As argued in \cite{Saad:2018bqo,Marolf:2020xie}, the connected part 
$\bra \prod_{i=1}^n Z(\bt_i)\ket_{\text{conn}}$ of this correlator comes from the contribution
of the Euclidean wormhole connecting the $n$ boundaries.

Of particular interest is the two-point function
$\bra Z(\bt_1)Z(\bt_2)\ket$ or its analytic continuation 
$\bra Z(\bt+\ri t)Z(\bt-\ri t)\ket$, known as the spectral form factor. 
The spectral form factor is a useful diagnostic of 
the quantum chaos \cite{Cotler:2016fpe,Garcia-Garcia:2016mno} and
exhibits the characteristic behavior called
ramp and plateau, as a function of time. The ramp comes from the eigenvalue
correlations \cite{Hikami} while the plateau arises from the pair-creation of eigenvalue
instantons \cite{Okuyama:2018gfr}.
The transition from the ramp to plateau occurs at what is called
Heisenberg time $t_H\sim \gs^{-1}$, where $\gs$ is the genus-counting parameter.
Around this time scale, the operator insertion $Z(\bt\pm\ri t)$ into the matrix integral
back-reacts to the eigenvalue distribution and two eigenvalues are pulled out from 
the dominant support (or cut) of the eigenvalue distribution.

This reminds us of the ``giant Wilson loop'' in 
4d $\mathcal{N}=4$ $\mathrm{SU}(N)$ super Yang--Mills
theory.\footnote{We would like thank Shota Komatsu for the discussion
on the analogy with giant Wilson loops.}
In that case, the path integral of the expectation value of the $1/2$ Bogomol'nyi--Prasad--Sommerfield Wilson loops
reduces to the Gaussian matrix integral
\cite{Erickson:2000af,Pestun:2007rz}.
For the winding Wilson loop with winding number $k$, when $k$ is of the
order of $N^0$,
the dual object is a fundamental string
on $AdS_5\times S^5$, but when $k$ becomes of the order of $N$ the bulk dual object morphs into
a D3-brane \cite{Drukker:2005kx}.
In the Gaussian matrix model picture,
what is happening for $k\sim\cO(N)$ 
is that one eigenvalue is pulled out from the cut 
due to the insertion of the large Wilson loop operator into the matrix integral
\cite{Kawamoto:2008gp,Hartnoll:2006is}. 
This is exactly the same mechanism as the ramp--plateau transition 
in the spectral form factor \cite{Okuyama:2018gfr}.
The different bulk dual pictures of the winding Wilson loop
for $k\sim\cO(N^0)$ and $k\sim\cO(N)$
are reflected in the different 
forms of the genus expansion: closed string like expansion for
$k\sim\cO(N^0)$ and open string like expansion for $k\sim\cO(N)$.

The above discussion suggests that one can study the open string like
expansion of the correlators $\bra \prod_i Z(\bt_i)\ket$
by taking the following scaling limit
\begin{align}
\label{eq:sregime}
\gs\ll 1,\quad \beta_i\gg 1\quad\mbox{with}\quad
s_i=\frac{\gs\beta_i}{2}\quad\mbox{fixed},
\end{align}
which we call the 't Hooft limit. Indeed, in our previous papers
\cite{Okuyama:2019xbv,Okuyama:2020ncd} we studied the 't Hooft limit of
the multi-boundary correlators in the JT gravity matrix model.
In \cite{Okuyama:2019xbv} we pointed out that
JT gravity is a special case of general
Witten--Kontsevich topological gravity,
where infinitely many couplings $t_k$ are
turned on with a specific value $t_k=\gamma_k$ with
\begin{align}
\label{eq:JTvalue}
\gamma_0=\gamma_1=0,\quad
\gamma_k=\frac{(-1)^k}{(k-1)!}\quad (k\ge 2).
\end{align}
In this paper we consider the 't Hooft expansion of 
the correlators $\bra \prod_i Z(\bt_i)\ket$ for 
Witten--Kontsevich topological gravity
with general couplings $t_k$.
We find that the leading term of this expansion is closely related to the
open free energy defined via the Laplace transform of the Baker--Akhiezer function
\cite{Buryak:2015eza,Dijkgraaf:2018vnm}.
We also show that the higher-order corrections to the 't Hooft expansion of
the correlators $\bra \prod_i Z(\bt_i)\ket$ can be systematically obtained from the
Korteweg--De Vries (KdV) equation \eqref{eq:WnKdV} or \eqref{eq:bketeq}.
It turns out that the 't Hooft expansion of the one-point function 
is obtained by using a similar algorithm for the computation of open free energy
developed in \cite{Okuyama:2020vrh}, while the 't Hooft expansion of an $n$-point function
with $n\geq2$ is determined algebraically from the lower point functions.

This paper is organized as follows. In section \ref{sec:review},
we briefly review how to compute the genus expansion of
the multi-boundary correlators by solving the KdV equation.
In section \ref{sec:1pt},
we present a method of computing the 't Hooft expansion of
the one-point function.
In section \ref{sec:2pt},
we explain that the 't Hooft expansion of the two-point function
can be computed algebraically.
In section \ref{sec:general},
we show that the 't Hooft expansion of a general $n$-point function
can also be computed algebraically.
In section \ref{sec:3pt},
we use this method to obtain the 't Hooft expansion
of the three-point function. We also calculate the exact result
in the Airy case.
Finally, we conclude in section \ref{sec:conclusion}.
In appendix \ref{sec:proof}, we present a proof of the relation
\eqref{eq:bketeq}.

\section{Multi-boundary correlators in topological gravity}\label{sec:review}

In Witten--Kontsevich topological gravity
\cite{Witten:1990hr,Kontsevich:1992ti}
(see e.g.~\cite{Dijkgraaf:2018vnm} for a recent review)
observables are made up of the intersection numbers
\begin{align}
\label{eq:intsec}
\langle\tau_{d_1}\cdots\tau_{d_n}\rangle_{g,n}
=\int_{\overline{\cal M}_{g,n}}
 \psi_1^{d_1}\cdots\psi_n^{d_n},\qquad
d_1,\ldots,d_n\in\bbZ_{\ge 0}.
\end{align}
They are associated with a closed Riemann surface $\Sigma$ of genus $g$
with $n$ marked points $p_1,\ldots,p_n$. We let ${\cal M}_{g,n}$
denote the moduli space of $\Sigma$ and $\overline{\cal M}_{g,n}$
the Deligne--Mumford compactification of ${\cal M}_{g,n}$.
Here $\tau_{d_i}=\psi_i^{d_i}$ and $\psi_i$ is the first Chern class of
the complex line bundle whose fiber is the cotangent space to $p_i$.
The generating function for the intersection numbers is defined as
\begin{align}
\label{eq:genF}
F(\{t_k\}):=\sum_{g=0}^\infty \gs^{2g-2}F_g(\{t_k\}),\qquad
F_g(\{t_k\})
 :=\left\langle e^{\sum_{d=0}^\infty t_d\tau_d}\right\rangle_g.
\end{align}

In this paper we consider the $n$-boundary connected correlator
(which we also call the $n$-point function)
\cite{Moore:1991ir}
\begin{align}
\label{eq:ZninF}
Z_n(\{\beta_i\},\{t_k\})
 =\left\bra \prod_{i=1}^nZ(\bt_i)\right\ket_{\text{conn}}
 \simeq B(\beta_1)\cdots B(\beta_n)F(\{t_k\}),
\end{align}
where
\begin{align}
\label{eq:Bdef}
B(\beta)
 :=\gs\sqrt{\frac{\beta}{2\pi}}\sum_{d=0}^\infty\beta^d
\frac{\partial}{\partial t_d}.
\end{align}
$B(\beta)$ can be thought of as the ``boundary creation operator.''
The symbol ``$\simeq$'' in \eqref{eq:ZninF} means that
the equality holds up to
an additional non-universal part \cite{Moore:1991ir}
when $3g-3+n<0$. Such a deviation appears only
in the genus-zero part of $n=1$-, 2-boundary correlators
and does not affect their higher genus parts
nor correlators with $n\ge 3$
(see e.g.~\cite{Okuyama:2020ncd} for a more detailed explanation).

$Z_n$ as well as $F$ satisfy a set of simple differential equations,
which allows us to compute their genus expansion.
To see this, let us first introduce the notation
\begin{align}
\label{eq:rescaledvar}
\hbar:=\frac{\gs}{\sqrt{2}},\quad
x :=\frac{t_0}{\hbar}\quad
\tau :=\frac{t_1}{\hbar}
\end{align}
and
\begin{align}
\label{eq:rescaleddiff}
\partial_k :=\frac{\partial}{\partial t_k},\quad
{}^{'}:=\partial_x=\hbar\partial_0,\quad
\dot{~}:=\partial_\tau =\hbar\partial_1.
\end{align}
The differential equations are simply written in terms of
the derivatives
\begin{align}
W_n:=Z_n',\qquad
W_0:=F',\qquad
u:=\gs^2\partial_0^2 F=2F''.
\label{eq:Wudef}
\end{align}
Recall that $u$ satisfies the KdV equation
\cite{Witten:1990hr,Kontsevich:1992ti}
\begin{align}
\label{eq:uKdV}
\dot{u}=uu'+\frac{1}{6}u'''.
\end{align}
Integrating this equation once in $x$ we obtain
\begin{align}
\label{eq:W0KdV}
\dot{W}_0
 &=(W_0')^2+\frac{1}{6}W_0'''.
\end{align}
Since $B(\beta_i)$ commutes with
$\dot{~}=\partial_\tau$ and ${}'=\partial_x$,
we immediately obtain a differential equation for
$W_n\simeq B(\beta_1)\cdots B(\beta_n)W_0$
by simply applying $B(\beta_1)\cdots B(\beta_n)$
to both sides of the above equation.
The result is \cite{Okuyama:2020ncd}
\begin{align}
\label{eq:WnKdV}
\begin{aligned}
\dot{W}_n(\beta_1,\ldots,\beta_n)
 &=\sum_{I\subset N}W_{|I|}'W_{|N-I|}'
   +\frac{1}{6}W_n'''(\beta_1,\ldots,\beta_n).
\end{aligned}
\end{align}
Here $N=\{1,2,\ldots,n\}$,
$W_{|I|}'=W_{|I|}'(\beta_{i_1},\ldots,\beta_{i_{|I|}})$
with $I=\{i_1,i_2,\ldots,i_{|I|}\}$,
and the sum is taken for all possible subsets $I$ of $N$
including the empty set.

As explained in \cite{Okuyama:2020ncd}
one can solve this equation and compute the genus expansion of
$W_n$ up to any order.
The genus expansion of $Z_n$ is then obtained by merely integrating
$W_n$ once in $x$. This can be done without ambiguity.
In \cite{Okuyama:2020ncd}
we demonstrated this computation in the (off-shell) JT gravity case
$t_k=\gamma_k\ (k\ge 2)$,
but as detailed in \cite{Okuyama:2020qpm},
all the results are immediately
generalized to the case of general $t_k$ by merely replacing
\begin{align}
\label{BIrel}
B_n\to (-1)^{n+1}I_{n+1}\qquad (n\ge 1).
\end{align}
Here
\begin{align}
\label{eq:defIn}
I_n=
I_n(u_0,\{t_k\})=\sum_{\ell=0}^\infty t_{n+\ell}\frac{u_0^\ell}{\ell!}
\quad (n\ge 0)
\end{align}
are Itzykson--Zuber variables \cite{Itzykson:1992ya},
\begin{align}
\label{eq:defu0}
u_0&:=\partial_0^2 F_0
\end{align}
is the genus-zero part of $u$, and
$B_n$ are Itzykson--Zuber variables restricted to the JT gravity case
\begin{align}
\begin{aligned}
B_n
 &=(-1)^{n+1}I_{n+1}\left(u_0,\left\{
   t_k=\gamma_k\ (k\ge 2)\right\}\right)\quad (n\ge 1)\\
 &=\sum_{k=0}^\infty\frac{(-1)^k u_0^k}{k!(k+n)!}.
\end{aligned}
\end{align}
The key to solving \eqref{eq:WnKdV} order by order
is the change of variables\footnote{
This change of variables was originally introduced by Zograf
(see e.g.~\cite{Zograf:2008wbe}).}
\begin{align}
\label{eq:d01inyt}
\partial_0
 =\frac{1}{t}(\partial_{u_0}-I_2\partial_t),
\qquad
\partial_1=u_0\partial_0-\partial_t.
\end{align}
That is, 
instead of $t_0$ and $t_1$ we take $u_0$ and
\begin{align}
\label{eq:tdef}
t:=(\partial_0 u_0)^{-1}=1-I_1
\end{align}
as independent variables and regard $t_{k\ge 2}$ as parameters.
In the new variables the integration constant is trivially fixed
at every step of solving the differential equation.
This is ensured by $F_1=-\frac{1}{24}\log t$ and by the fact that
$F_g\ (g\ge 2)$ are polynomials
in the generators $I_{n\ge 2}$ and $t^{-1}$
\cite{Itzykson:1992ya,Eguchi:1994cx,Zhang:2019hly}.

To summarize, we know that one can compute the small $\gs$ expansion of
the $n$-boundary correlator $Z_n$ up to any order.
This expansion can be thought of as a closed string like expansion.
Interestingly, $Z_n$ also admits an open string like expansion.
This is again a small $\gs$ expansion,
but is performed in the 't Hooft regime \eqref{eq:sregime}.
In the rest of the paper we will show that
one can also compute this expansion up to any order.

\section{'t Hooft expansion of the one-boundary correlator}\label{sec:1pt}

In the scaling regime \eqref{eq:sregime},
the one-point function admits the 't Hooft expansion\footnote{
$\cF_\dg$ in this paper are related to those in our previous work
\cite{Okuyama:2019xbv,Okuyama:2020ncd} by 
$\cF_\dg^{\rm here}
 \leftrightarrow\sqrt{2}^{1-\dg}\cF_\dg^{\rm there}$
with the identification $\lambda=\sqrt{2}s$.}
\begin{align}
\label{eq:cFexp}
\cF = \log Z_1 = \sum_{\dg=0}^\infty\gs^{\dg-1}\cF_\dg,
\end{align}
where $\cF_\dg$ is a function of $s=\gs\bt/2$ and $t_k$.
In \cite{Okuyama:2019xbv}
we calculated $\cF_\dg$ with $\dg=0,1,2$
in the JT gravity case $t_k=\gamma_k\ (k\ge 0)$
by saddle point method.
In \cite{Okuyama:2020ncd} we generalized the calculation to
the ``off-shell'' case $t_k=\gamma_k\ (k\ge 2)$ with $t_0,t_1$
being unfixed.
In what follows we will present a method of
computing $\cF_\dg$ up to any $\dg$ with general $t_k$
by solving the differential equation \eqref{eq:WnKdV}.
The method is very similar to that of computing
the genus expansion of the open free energy \cite{Okuyama:2020vrh}
and is much more efficient than the saddle point calculation.

Instead of directly dealing with \eqref{eq:cFexp},
we first compute the genus expansion
\begin{align}
\label{eq:Gexp}
G=\log W_1=\sum_{\dg=0}^\infty \gs^{\dg-1} G_\dg.
\end{align}
Since $G$ and $\cF$ are related by
\begin{align}
\label{eq:GFrel}
G=\cF+\log\partial_x\cF,
\end{align}
the expansion \eqref{eq:cFexp} will immediately be obtained
once \eqref{eq:Gexp} is computed.
The differential equation \eqref{eq:WnKdV} for $n=1$ is written as
\begin{align}
\label{eq:W1KdV}
\partial_1 W_1 = u\partial_0 W_1 +\frac{\gs^2}{12}\partial_0^3 W_1.
\end{align}
This implies
\begin{align}
\label{eq:Gdiffeq}
\partial_1 G
 =u\partial_0 G
 +\frac{\gs^2}{12}\Bigl(\partial_0^3 G+3\partial_0 G\partial_0^2 G
   +(\partial_0 G)^3 \Bigr),
\end{align}
which is rewritten as
\begin{align}
\label{eq:Gdiffeq2}
-\partial_t G
 =\frac{\gs^2}{12}(\partial_0 G)^3
 +\frac{\gs^2}{4}\partial_0 G\partial_0^2 G
 +\frac{\gs^2}{12}\partial_0^3 G
 +(u-u_0)\partial_0 G.
\end{align}
Recall that the genus expansion
\begin{align}
\label{eq:uexp}
u&=\sum_{g=0}^\infty \gs^{2g}u_g
\end{align}
is computed by solving the KdV equation \eqref{eq:uKdV}
(see e.g.~\cite{Okuyama:2019xbv,Okuyama:2020vrh}
for the results in our convention).
By plugging \eqref{eq:Gexp} and \eqref{eq:uexp} into
\eqref{eq:Gdiffeq2}, one obtains
\begin{align}
\label{eq:Grec01}
\begin{aligned}
-\partial_t G_0&=\frac{1}{12}(\partial_0 G_0)^3,\\
\nD G_1&=\frac{1}{4}\partial_0 G_0\partial_0^2 G_0
\end{aligned}
\end{align}
for $\dg=0,1$ and
\begin{align}
\label{eq:Grecrel}
\begin{aligned}
\nD G_\dg
  =\frac{1}{12}\sum_{\substack{0\le i,j,k<\dg\\[1ex] i+j+k=\dg}}
     \partial_0 G_i \partial_0 G_j \partial_0 G_k
  +\frac{1}{4}\sum_{k=0}^{\dg-1}\partial_0 G_{\dg-k-1} \partial_0^2 G_{k}
  +\frac{1}{12}\partial_0^3 G_{\dg-2}
  +\sum_{k=1}^{\lfloor\frac{\dg}{2}\rfloor}u_k\partial_0 G_{\dg-2k}
\end{aligned}
\end{align}
for $\dg\ge 2$.
Here we have introduced the differential operator
\begin{align}
\label{eq:nDdef}
\nD := -\partial_t-\frac{1}{4}(\partial_0 G_0)^2\partial_0.
\end{align}

In what follows we will solve the above differential equations
and compute $G_\dg$.
First of all, the explicit form of $G_0$ is obtained as follows.
Recall that 
$W_1$ is related to the Baker--Akhiezer function $\psi(\xi)$
as \cite{Okuyama:2019xbv}
\begin{align}
W_1(s)
 =e^{G(s)}
 =\int_{-\infty}^\infty d\xi e^{\frac{2s\xi}{\gs}}\psi(\xi)^2
\label{eq:W1inpsi}
\end{align}
and $\psi(\xi)$ is expanded as
\cite{Okuyama:2019xbv,Okuyama:2020vrh}
\begin{align}
\label{eq:psiexp}
\psi(\xi)
 =\exp\Biggl(\sum_{\dg=0}^\infty \gs^{\dg-1}A_\dg\Biggr)
\end{align}
with
\begin{align}
A_0=-\frac{tz^3}{3}+\sum_{n=1}^\infty\frac{I_{n+1}}{(2n+3)!!}z^{2n+3},
\qquad z=\sqrt{2(\xi-u_0)}.
\end{align}
The integral \eqref{eq:W1inpsi}
can be evaluated by the saddle point method. $G_0$ is given by
\begin{align}
\label{eq:G0saddle}
G_0=2s\xi_* +2A_0(\xi_*),
\end{align}
where the saddle point $\xi_*$ is determined by the condition
\begin{align}
\label{eq:saddlecond}
\partial_\xi\left[2s\xi+2A_0(\xi)\right]\Big|_{\xi=\xi_*}=0.
\end{align}
This is equivalent to
\begin{align}
\label{eq:sinzs}
s=-\partial_\xi A_0\Big|_{\xi=\xi_*}
 =t\zs-\sum_{n=1}^\infty\frac{I_{n+1}}{(2n+1)!!}\zs^{2n+1},
\qquad \zs=\sqrt{2(\xi_*-u_0)}.
\end{align}
As we showed in \cite{Okuyama:2020vrh},
this relation is inverted as
\begin{align}
\label{eq:zsins}
\zs
 &=\sum_{\substack{j_a\ge 0\\[.5ex]
                   \sum_a j_a=k\\[.5ex]
                   \sum_a aj_a=n}}
 \frac{(2n+k)!}{(2n+1)!}\frac{s^{2n+1}}{t^{2n+k+1}}
 \prod_{a=1}^\infty\frac{I_{a+1}^{j_a}}{j_a!(2a+1)!!^{j_a}}.
\end{align}
Plugging this back into \eqref{eq:G0saddle},
we obtain the explicit form of $G_0$.
In fact, $G_0$ is exactly twice
the genus zero part $\Fo_0$ of the open free energy
studied in \cite{Okuyama:2020vrh}, for which
the following simple expression is available:
\begin{align}
\label{eq:Fo0}
G_0=2\Fo_0
&=2u_0s
 +2\sum_{\substack{j_a\ge 0\\[.5ex]
                   \sum_a j_a=k\\[.5ex]
                   \sum_a aj_a=n}}
 \frac{(2n+k+1)!}{(2n+3)!}\frac{s^{2n+3}}{t^{2n+k+2}}
 \prod_{a=1}^\infty\frac{I_{a+1}^{j_a}}{j_a!(2a+1)!!^{j_a}}.
\end{align}
It also follows that \cite{Okuyama:2020vrh}
\begin{align}
\label{eq:d0G0}
\partial_0 G_0= 2\zs,\qquad
\partial_s G_0= 2\xi_*.
\end{align}
In terms of $z_*$, the differential equations \eqref{eq:Grec01}
are written as
\begin{align}
\label{eq:Grec01bis}
\begin{aligned}
-\partial_t G_0 &= \frac{2}{3}\zs^3,\\
\nD G_1 &= \zs\partial_0\zs
\end{aligned}
\end{align}
and $D$ in \eqref{eq:nDdef} becomes
\begin{align}
\label{eq:nD}
\nD=-\partial_t-z_*^2\partial_0.
\end{align}
We saw in \cite{Okuyama:2020vrh} that $G_0=2\Fo_0$ indeed satisfies
the first equation in \eqref{eq:Grec01bis}.

The operator $\nD$ has interesting properties.
(This is analogous to $\cD$ in \cite{Okuyama:2020vrh}.)
For instance, we see that
\begin{align}
\begin{aligned}
\nD\zs
 &=-\partial_t\zs -\zs^2\partial_0\zs
  =\zs\partial_s\zs-\zs\partial_0(\xi_*-u_0)
  =\zs\partial_0\xi_*-\zs\partial_0(\xi_*-u_0)\\
 &=\frac{\zs}{t},\\
\nD\xi_*
 &=-\partial_t\xi_* -\zs^2\partial_0\xi_*
  =-\zs\partial_t\zs-\zs^2\partial_s\zs\\
 &= 0,
\end{aligned}
\end{align}
where we have used \cite{Okuyama:2020vrh}
\begin{align}
-\partial_t\zs=\zs\partial_s\zs.
\end{align}
We can also show that
\begin{align}
\begin{aligned}
\nD\xis{n}
 &=\nD\partial_s\xis{n-1}\\
 &=-\partial_t\partial_s\xis{n-1}-\zs^2\partial_0\partial_s\xis{n-1}\\
 &=-\partial_s\partial_t\xis{n-1}
   -\partial_s\bigl(\zs^2\partial_0\xis{n-1}\bigr)
   +(\partial_s\zs^2)\partial_0\xis{n-1}\\
 &=\partial_s\nD\xis{n-1}
   +2\zs(\partial_s\zs)\partial_s^{n-1}\partial_0\xi_*\\
 &=\partial_s\nD\xis{n-1}
   -2(\partial_t\zs)\zs^{(n)},
\end{aligned}
\end{align}
where
\begin{align}
\xis{n}:=\partial_s^n\xi_*,\qquad
\zs^{(n)}:=\partial_s^n\zs\qquad (n\ge 1).
\end{align}
From this we find
\begin{align}
\begin{aligned}
\nD\xis{n}
 &=-2\sum_{k=0}^{n-1}\partial_s^k\bigl(\zs^{(n-k)}\partial_t\zs\bigr)\\
 &=-2\sum_{k=0}^{n-1}\sum_{\ell=0}^k
   \left(\begin{array}{@{}c@{}} k\\[1ex] \ell \end{array}\right)
   \bigl(\partial_s^{k-\ell}\zs^{(n-k)}\bigr)
   \partial_s^{\ell}\partial_t\zs\\
 &=2\sum_{k=0}^{n-1}\sum_{\ell=0}^k
   \left(\begin{array}{@{}c@{}} k\\[1ex] \ell \end{array}\right)
   \zs^{(n-\ell)}\partial_s^{\ell}(\zs\partial_s\zs)\\
 &=2\sum_{k=0}^{n-1}\sum_{\ell=0}^k
   \left(\begin{array}{@{}c@{}} k\\[1ex] \ell \end{array}\right)
   \zs^{(n-\ell)}
   \sum_{m=0}^\ell
   \left(\begin{array}{@{}c@{}} \ell\\[1ex] m \end{array}\right)
   \zs^{(\ell-m)}\zs^{(m+1)}\\
 &=2\sum_{k=0}^{n-1}\sum_{\ell=0}^k\sum_{m=0}^\ell
   \frac{k!}{(k-\ell)!(\ell-m)!m!}
   \zs^{(n-\ell)}\zs^{(\ell-m)}\zs^{(m+1)}.
\end{aligned}
\end{align}
As explained in \cite{Okuyama:2020vrh},
$\zs^{(n\ge 1)}$ can be expressed in terms of $\xis{n\ge 1}$ and $\zs$:
\begin{align}
\label{eq:zinxi}
\begin{aligned}
\zs^{(1)}&=\frac{\xi_*^{(1)}}{\zs},\\
\zs^{(2)}
 &=\frac{\xi_*^{(2)}}{\zs}
  -\frac{\bigl(\xi_*^{(1)}\bigr)^2}{\zs^3},\\
\zs^{(3)}
 &=\frac{\xi_*^{(3)}}{\zs}
  -\frac{3\xi_*^{(1)}\xi_*^{(2)}}{\zs^3}
  +\frac{3\bigl(\xi_*^{(1)}\bigr)^3}{\zs^5}.
\end{aligned}
\end{align}
Therefore $\nD\xis{n\ge 1}$ can also be expressed in terms of
$\xis{n\ge 1}$ and $\zs$. For $n=1,2,3$ we have
\begin{align}
\begin{aligned}
\nD\xis{1}
 &=\frac{2\bigl(\xis{1}\bigr)^2}{\zs},\\
\nD\xis{2}
 &=\frac{6\xis{1}\xis{2}}{\zs}-\frac{4\bigl(\xis{1}\bigr)^3}{\zs^3},\\
\nD\xis{3}
 &=\frac{8\xis{1}\xis{3}}{\zs}+\frac{6\bigl(\xis{2}\bigr)^2}{\zs}
  -\frac{24\bigl(\xis{1}\bigr)^2\xis{2}}{\zs^3}
  +\frac{18\bigl(\xis{1}\bigr)^4}{\zs^5}.
\end{aligned}
\end{align}
On the other hand, as in \cite{Okuyama:2020vrh}
we evaluate the r.h.s.~of \eqref{eq:Grecrel} using
\begin{align}
\label{eq:d0tzs}
\begin{aligned}
\partial_0\zs&=\frac{\zs^{(1)}}{\zs}-\frac{1}{t\zs},\\
\partial_0\xi_*^{(n)}&=\zs^{(n+1)}\quad (n\ge 0).
\end{aligned}
\end{align}
In this way, one can express both sides of \eqref{eq:Grecrel}
as a polynomial in the variables 
$t^{-1}$, $I_{k\ge 2}$, $\zs^{-1}$, $\bigl(\xis{1}\bigr)^{-1}$
and $\xis{n\ge 1}$.

Almost in the same way as in \cite{Okuyama:2020vrh},
we can formulate the following
algorithm to solve \eqref{eq:Grecrel}
and obtain $G_{\dg}$ from
the data of $\{G_{\dg'}\}_{\dg'<\dg}$:
\renewcommand{\theenumi}{\roman{enumi}}
\renewcommand{\labelenumi}{(\theenumi)}
\begin{enumerate}
\item Compute the r.h.s.~of \eqref{eq:Grecrel}
and express it as a polynomial in the variables 
$t^{-1}$, $I_{k\ge 2}$, $\zs^{-1}$, $\bigl(\xis{1}\bigr)^{-1}$
and $\xis{n\ge 1}$.

\item Let $t^{-m}f(I_k,\zs,\xi_*^{(n)})$
denote the highest-order part in $t^{-1}$
of the obtained expression. This part can arise only from
\begin{align}
\nD\left(-\frac{f(I_k,\zs,\xi_*^{(n)})}{(m-2)t^{m-2}\zs^2 I_2}\right).
\end{align}
Therefore subtract this from the obtained expression.

\item Repeat procedure (ii) down to $m=3$.
Then all the terms of order $t^{-2}$ automatically disappear
and the remaining terms are of order $t^{-1}$ or $t^{0}$.
Note also that the expression does not contain any $I_k$.

\item In the result of (iii),
collect all the terms of order $t^{-1}$
and let $t^{-1}\zs\partial_{\zs}g(\zs,\xi_*^{(n)})$
denote the sum of them. This part arises from
\begin{align}
\nD g(\zs,\xi_*^{(n)}).
\end{align}
Therefore subtract this from the result of (iii).
The remainder turns out to be independent of $t$.

\item
In the obtained expression,
let
\begin{align}
\frac{h\bigl(\xis{n\ge 2}\bigr)}{\zs\bigl(\xis{1}\bigr)^m}
\end{align}
denote the part which is of the order $\zs^{-1}$
as well as of the lowest order in $\bigl(\xis{1}\bigr)^{-1}$.
This part arises from
\begin{align}
\nD\left(
\frac{h\bigl(\xis{n\ge 2}\bigr)}{2(m+1)\bigl(\xis{1}\bigr)^{m+1}}\right).
\end{align}
Therefore subtract this from the obtained expression.

\item Repeat procedure (v) until the resulting expression vanishes.

\item By summing up all the above-obtained
primitive functions, we obtain $G_\dg$.

\end{enumerate}
Using this algorithm we can compute $G_\dg$ up to a high order.
($G_1$ is also obtained by solving \eqref{eq:Grec01bis}.)
The first few $G_\dg$ terms are
\begin{align}
\begin{aligned}
G_1&=\frac{1}{2}\log\xis{1}-\log\zs,\\
G_2&=
-\frac{I_2}{12t^2\zs}
-\frac{5}{12t\zs^3}
+\frac{3\xis{1}}{4\zs^4}
-\frac{\xis{2}}{4\zs^2\xis{1}}
+\frac{\xis{3}}{16\bigl(\xis{1}\bigr)^2}
-\frac{\bigl(\xis{2}\bigr)^2}{12\bigl(\xis{1}\bigr)^3},\\
G_3&=
\frac{I_2^2}{12t^4\zs^2}
+\frac{I_3}{24t^3\zs^2}
+\frac{I_2}{4t^3\zs^4}
+\frac{5}{8t^2\zs^6}
-\frac{5I_2\xis{1}}{48t^2\zs^5}
+\frac{I_2\xis{2}}{48t^2\zs^3\xis{1}}
+\frac{5\xis{2}}{16t\zs^5\xis{1}}
-\frac{35\xis{1}}{16t\zs^7}\\
&\hspace{1em}
-\frac{11\xis{2}}{8\zs^6}
-\frac{\xis{4}}{32\zs^2\bigl(\xis{1}\bigr)^2}
+\frac{7\xis{3}\xis{2}}{48\zs^2\bigl(\xis{1}\bigr)^3}
-\frac{\bigl(\xis{2}\bigr)^3}{8\zs^2\bigl(\xis{1}\bigr)^4}
+\frac{3\bigl(\xis{1}\bigr)^2}{\zs^8}
+\frac{3\xis{3}}{16\zs^4\xis{1}}\\
&\hspace{1em}
-\frac{\bigl(\xis{2}\bigr)^2}{8\zs^4\bigl(\xis{1}\bigr)^2}
+\frac{\xis{5}}{192\bigl(\xis{1}\bigr)^3}
-\frac{\bigl(\xis{3}\bigr)^2}{32\bigl(\xis{1}\bigr)^4}
-\frac{\xis{4}\xis{2}}{24\bigl(\xis{1}\bigr)^4}
+\frac{3\xis{3}\bigl(\xis{2}\bigr)^2}{16\bigl(\xis{1}\bigr)^5}
-\frac{\bigl(\xis{2}\bigr)^4}{8\bigl(\xis{1}\bigr)^6}.
\end{aligned}
\end{align}

$\cF_\dg$ can easily be obtained
by inverting the relation \eqref{eq:GFrel}.
We obtain
\begin{align}
\label{eq:cFn}
\begin{aligned}
\cF_0
 &=G_0=2\Fo_0,\\
\cF_1
 &=G_1-\log(\hbar\partial_0\cF_0)\\
 &=\frac{1}{2}\log\xis{1}-2\log\zs-\log\hbar,\\
\cF_2
 &=G_2-\frac{\partial_0\cF_1}{\partial_0\cF_0}\\
 &=
-\frac{I_2}{12t^2\zs}
-\frac{17}{12t\zs^3}
+\frac{2\xis{1}}{\zs^4}
-\frac{\xis{2}}{2\zs^2\xis{1}}
+\frac{\xis{3}}{16\bigl(\xis{1}\bigr)^2}
-\frac{\bigl(\xis{2}\bigr)^2}{12\bigl(\xis{1}\bigr)^3},\\
\cF_3
 &=G_3-\frac{\partial_0\cF_2}{\partial_0\cF_0}
      +\frac{1}{2}
       \left(\frac{\partial_0\cF_1}{\partial_0\cF_0}\right)^2\\
 &=
\frac{I_2^2}{6t^4\zs^2}
+\frac{I_3}{12t^3\zs^2}
+\frac{I_2}{t^3\zs^4}
+\frac{13}{4t^2\zs^6}
-\frac{7I_2\xis{1}}{48t^2\zs^5}
+\frac{I_2\xis{2}}{48t^2\zs^3\xis{1}}
+\frac{17\xis{2}}{16t\zs^5\xis{1}}
-\frac{153\xis{1}}{16t\zs^7}\\
&\hspace{1em}
+\frac{7\xis{3}\xis{2}}{24\zs^2\bigl(\xis{1}\bigr)^3}
-\frac{3\bigl(\xis{2}\bigr)^2}{8\zs^4\bigl(\xis{1}\bigr)^2}
-\frac{\bigl(\xis{2}\bigr)^3}{4\zs^2\bigl(\xis{1}\bigr)^4}
+\frac{\xis{3}}{2\zs^4\xis{1}}
-\frac{\xis{4}}{16\zs^2\bigl(\xis{1}\bigr)^2}
+\frac{3\xis{3}\bigl(\xis{2}\bigr)^2}{16\bigl(\xis{1}\bigr)^5}\\
&\hspace{1em}
-\frac{\xis{4}\xis{2}}{24\bigl(\xis{1}\bigr)^4}
+\frac{10\bigl(\xis{1}\bigr)^2}{\zs^8}
-\frac{4\xis{2}}{\zs^6}
+\frac{\xis{5}}{192\bigl(\xis{1}\bigr)^3}
-\frac{\bigl(\xis{3}\bigr)^2}{32\bigl(\xis{1}\bigr)^4}
-\frac{\bigl(\xis{2}\bigr)^4}{8\bigl(\xis{1}\bigr)^6}.
\end{aligned}
\end{align}
We computed $\cF_\dg$ for $\dg\le 13$.
We have checked that the above $\cF_\dg$ with $\dg=0,1,2$
are in perfect agreement (up to the constant part of $\cF_1$)
with $\sqrt{2}^{1-\dg}\cF_\dg$
in \cite{Okuyama:2020ncd}
under the identification in \eqref{BIrel}.

\section{'t Hooft expansion of the two-boundary correlator}\label{sec:2pt}

In this section let us consider 't Hooft expansion of
the two-boundary correlator.
While $Z_2$ itself admits 't Hooft expansion,
for many purposes it is convenient to consider instead
the 't Hooft expansion of
\begin{align}
\tZ_2(s_1,s_2)
 =\Tr\left[e^{\beta_1 Q}\Pi e^{\beta_2 Q}\Pi\right]
 =Z_1(s_1+s_2)-Z_2(s_1,s_2),
\end{align}
where
\begin{align}
Q:=\partial_x^2+u,\qquad
\Pi :=\int_{-\infty}^x dx'|x'\rangle\langle x'|.
\end{align}
Correspondingly, let us define the derivative
\begin{align}
\tW_2(s_1,s_2)
  =\partial_x \tZ_2(s_1,s_2)
  =W_1(s_1+s_2)-W_2(s_1,s_2)
\end{align}
and ``free energies''
\begin{align}
\cK^{(2)}=\log \tZ_2,\qquad
G^{(2)}=\log \tW_2.
\end{align}
$G^{(2)}$ and $\cK^{(2)}$ are related by
\begin{align}
\label{eq:G2K2rel}
G^{(2)}=\cK^{(2)}+\log\partial_x\cK^{(2)}.
\end{align}
We consider the expansions
\begin{align}
\label{eq:G2K2exp}
G^{(2)}=\sum_{\dg=0}^\infty \gs^{\dg-1} G^{(2)}_\dg,\qquad
\cK^{(2)}=\sum_{\dg=0}^\infty \gs^{\dg-1} \cK^{(2)}_\dg,
\end{align}
where $G^{(2)}_\dg$ and $\cK^{(2)}_\dg$ are functions of
$s_i=\gs\bt_i/2~(i=1,2)$ and $t_k$.

The differential equation \eqref{eq:WnKdV} for $n=2$
is written as
\begin{align}
\label{eq:W2KdV}
\partial_1 W_2
 =u\partial_0 W_2
  +\frac{\gs^2}{12}\partial_0^3 W_2
  +\sqrt{2}\gs\partial_0W_1(s_1)\partial_0W_1(s_2).
\end{align}
Subtracting \eqref{eq:W2KdV} from
\eqref{eq:W1KdV} with $s=s_1+s_2$, we obtain
the differential equation for $\tW_2(s_1,s_2)$:
\begin{align}
\label{eq:tW2KdV}
-\partial_t\tW_2
 =(u-u_0)\partial_0\tW_2
  +\frac{\gs^2}{12}\partial_0^3 \tW_2
  -\sqrt{2}\gs\partial_0W_1(s_1)\partial_0W_1(s_2).
\end{align}
This implies
\begin{align}
\label{eq:G2KdV}
\begin{aligned}
&\hspace{-1em}
\sqrt{2}\gs\partial_0 G(s_1)\partial_0 G(s_2)
e^{G(s_1)+G(s_2)-G^{(2)}}\\
 &=\partial_t G^{(2)}
  +(u-u_0)\partial_0 G^{(2)}
  +\frac{\gs^2}{12}\Bigl(
     \partial_0^3 G^{(2)}+3\partial_0 G^{(2)}\partial_0^2 G^{(2)}
    +(\partial_0 G^{(2)})^3 \Bigr).
\end{aligned}
\end{align}
In \cite{Okuyama:2020ncd} we derived that
\begin{align}
\label{eq:K2init}
\cK^{(2)}_0(s_1,s_2)=\cF_0(s_1)+\cF_0(s_2).
\end{align}
From this and \eqref{eq:G2K2rel} we have
\begin{align}
\label{eq:G2init}
G^{(2)}_0(s_1,s_2)=G_0(s_1)+G_0(s_2)
\end{align}
as the initial condition.

We observe that starting with \eqref{eq:G2init}
and comparing both sides of \eqref{eq:G2KdV}
order-by-order in the small $\gs$ expansion
one can algebraically determine $G^{(2)}_\dg$
from the data of $\bigl\{G^{(2)}_{\dg'}\bigr\}_{\dg'<\dg}$
and $\{G_{\dg'}\}_{\dg'\le\dg}$.
For instance, for $\dg=1$ we obtain
\begin{align}
\begin{aligned}
G^{(2)}_1
 &=\frac{1}{2}\log\xi_{1*}^{(1)}
  +\frac{1}{2}\log\xi_{2*}^{(1)}
  -\log z_{1*}
  -\log z_{2*}
  -\log(z_{1*}+z_{2*})+\frac{3}{2}\log 2.
\end{aligned}
\end{align}
As in the case of one-point function,
$\cK^{(2)}_\dg$ can also be algebraically
determined by \eqref{eq:G2K2rel} from
the data of $\bigl\{G^{(2)}_{\dg'}\bigr\}_{\dg'\le\dg}$.
Therefore, given the data of $\{G_{\dg'}\}_{\dg'\le\dg}$
we can compute $\cK^{(2)}_\dg$ without
any integration procedure. For instance, we obtain
\begin{align}
\begin{aligned}
\cK^{(2)}_1
 &=G^{(2)}_1-\log\bigl(\hbar\partial_0\cK^{(2)}_0\bigr)\\
 &=\frac{1}{2}\log\xi_{1*}^{(1)}
  +\frac{1}{2}\log\xi_{2*}^{(1)}
  -\log z_{1*}
  -\log z_{2*}
  -2\log(z_{1*}+z_{2*})+\frac{1}{2}\log 2
  -\log\hbar.
\end{aligned}
\end{align}

Using the above method we computed $\cK^{(2)}_\dg$ for $\dg\le 8$.
We verified that $\cK^{(2)}_\dg$ with $\dg=1,2$
are in perfect agreement (up to the constant part of $\cK^{(2)}_1$)
with $\sqrt{2}^{1-\dg}\cK^{(2)}_\dg$
given in (3.52) of \cite{Okuyama:2020ncd}.

As a further nontrivial check,
let us compare the above results with
the low-temperature expansion of the two-point function
studied in \cite{Okuyama:2020ncd}.
As we mentioned, the results in \cite{Okuyama:2020ncd}
are trivially generalized to the case of general $t_k$ by
the replacement \eqref{BIrel}.
Recall that the low-temperature expansion of
$e^{\cK^{(2)}}$ is written as (see (4.32) of \cite{Okuyama:2020ncd}
and notations therein)
\begin{align}
\begin{aligned}
e^{\cK^{(2)}}
&=\Tr(e^{\beta_1 Q}\Pi e^{\beta_2 Q}\Pi)\\
&=\Erfc(\sqrt{D_0})
  \frac{e^{\frac{h^2}{12t^2}+\frac{y}{T}}}{2\sqrt{\pi}h}
  \sum_{\ell=0}^\infty\frac{T^\ell}{\ell!}z_\ell
  -\cB\sum_{\ell=0}^\infty\frac{T^{\ell+1}}{\ell!}g_\ell\\
&=\frac{1}{2\sqrt{\pi}h}e^{\frac{u_0}{T}+\frac{h^2}{12t^2}-D_0}
  \left[
  e^{D_0}\Erfc(\sqrt{D_0})
  \sum_{\ell=0}^\infty\frac{T^\ell}{\ell!}z_\ell
  -2t\sqrt{\frac{D_0}{\pi}}
  \sum_{\ell=0}^\infty\frac{T^\ell}{\ell!}g_\ell
  \right].
\end{aligned} 
\label{eq:TexpK2}
\end{align}
Using the data of $z_\ell, g_\ell\ (0\le \ell\le 7)$
we verified that this expression indeed reproduces
the above-obtained $\cK^{(2)}_\dg$ in the form of small-$s$ expansion.
We performed the expansion of $\cK^{(2)}_\dg (2\le \dg\le 5)$
up to the order of $s^{3(6-\dg)}$
and observed perfect agreement.
Since small-$s$ expansion of $\cK^{(2)}_\dg (\dg\ge 2)$
starts at the order of $s^{3(1-\dg)}$,
this serves as a rather nontrivial check.

\section{General formalism for multi-boundary correlator}\label{sec:general}

In this section let us consider
't Hooft expansion of multi-boundary correlators.
As in the case of the two-boundary correlator,
it is convenient to consider
the 't Hooft expansion
\begin{align}
\cK^{(n)}
 =\log\tZ_n
 =\sum_{\dg=0}^\infty\gs^{\dg-1}\cK^{(n)}_\dg
\label{eq:Kn-exp}
\end{align}
with
\begin{align}
\label{eq:tZn}
\tZ_n(\beta_1,\ldots,\beta_n)
 =\Tr(e^{\beta_1 Q}\Pi \cdots e^{\beta_n Q}\Pi).
\end{align}
$\cK^{(n)}_\dg$ in \eqref{eq:Kn-exp} is a function of 
$s_i=\gs\bt_i/2~(i=1,\ldots,n)$ and $t_k$.
As we saw in \cite{Okuyama:2020ncd}, $Z_n$ and $\tZ_n$ are related as
\begin{align}
\begin{aligned}
Z_1(\beta)
 &=\Tr\left[e^{\beta Q}\Pi\right]\\
 &=\tZ_1(\beta),\\
Z_2(\beta_1,\beta_2)
 &=\Tr\left[e^{(\beta_1+\beta_2)Q}\Pi
   -e^{\beta_1 Q}\Pi e^{\beta_2 Q}\Pi\right]\\
 &=\tZ_1(\beta_1+\beta_2)-\tZ_2(\beta_1,\beta_2),\\
Z_3(\beta_1,\beta_2,\beta_3)
 &=\Tr\left[e^{(\beta_1+\beta_2+\beta_3)Q}\Pi
   +e^{\beta_1 Q}\Pi e^{\beta_2 Q}\Pi e^{\beta_3 Q}\Pi
   +e^{\beta_1 Q}\Pi e^{\beta_3 Q}\Pi e^{\beta_2 Q}\Pi\right.\\
 &\hspace{2.7em}\left.
   -e^{\beta_1 Q}\Pi e^{(\beta_2+\beta_3) Q}\Pi
   -e^{\beta_2 Q}\Pi e^{(\beta_3+\beta_1) Q}\Pi
   -e^{\beta_3 Q}\Pi e^{(\beta_1+\beta_2) Q}\Pi\right]\\
 &=\tZ_1(\beta_1+\beta_2+\beta_3)+\tZ_3(\beta_1,\beta_2,\beta_3)
  +\tZ_3(\beta_1,\beta_3,\beta_2)\\
 &\hspace{1em}
  -\tZ_2(\beta_1,\beta_2+\beta_3)
  -\tZ_2(\beta_2,\beta_3+\beta_1)-\tZ_2(\beta_3,\beta_1+\beta_2)
\end{aligned}
\label{eq:Zn-tZn}
\end{align}
for $n=1,2,3$. In general, the relation is given by the formula
\cite{Okuyama:2018aij}
\begin{align}
\begin{aligned}
Z_n(\beta_1,\ldots,\beta_n)
 &=\Tr\log
   \left(1+\left[-1+\prod_{i=1}^n(1+z_i e^{\beta_iQ})\right]\Pi\right)
   \Bigg|_{{\cal O}(z_1\cdots z_n)}\\
 &=\Tr\log
   \left(1+\sum_{k=1}^n\sum_{i_1<\cdots<i_k}z_{i_1}\cdots z_{i_k}
           e^{(\beta_{i_1}+\cdots+\beta_{i_k})Q}\Pi\right)
   \Bigg|_{{\cal O}(z_1\cdots z_n)}.
\end{aligned}
\end{align}

Let us next introduce
\begin{align}
\tW_n(\beta_1,\ldots,\beta_n)=\partial_x\tZ_n(\beta_1,\ldots,\beta_n).
\end{align}
One may expect that \eqref{eq:WnKdV} leads to a differential equation
for $\tW_n$ similar to \eqref{eq:tW2KdV},
which enables us to compute the 't Hooft expansion of $\log\tW_n$
in the same way as in the last section.
However, this is not the case.
This is because
$\tW_n(\beta_1,\ldots,\beta_n)$ are only cyclically
symmetric with respect to the variables $\beta_i$
and for $n\ge 3$ multiple $\tW_n$
with different orders of $\beta_i$
appear in the single differential equation \eqref{eq:WnKdV}.
Consequently, the differential equation is not determinative
for $\tW_{n\ge 3}$.

On the other hand, by taking a different approach
it is still possible to compute 't Hooft expansion of
$\tW_{n\ge 3}$.
Our new approach is based on the fact that
$\tZ_n$ is expressed in terms of the Baker--Akhiezer function
$\psi_i:=\psi(\xi_i;\{t_k\})$,
which satisfies
\begin{align}
\label{eq:auxlin}
L\psi_i=\xi_i\psi_i,\qquad \dot{\psi}_i=M\psi_i,
\end{align}
where
\begin{align}
L=Q=\partial_x^2+u,\qquad
M=\frac{2}{3}\partial_x^3+u\partial_x+\frac{1}{2}u'.
\end{align}
From \eqref{eq:auxlin} we are able to derive a new differential equation,
which is not for $\tW_n$ itself, but for its constituents,
as we will see below.

We first recall that \eqref{eq:tZn} is rewritten as \cite{Okuyama:2020ncd}
\begin{align}
\tZ_n(\beta_1,\beta_2,\ldots,\beta_n)
 =\int_{-\infty}^\infty d\xi_1\cdots\int_{-\infty}^\infty d\xi_n
  e^{\sum_{j=1}^n\beta_j\xi_j}
  K_{12}K_{23}\cdots K_{n-1,n}K_{n,1},
\end{align}
where $K_{ij}$ is the Darboux--Christoffel kernel.
It is written
in terms of the Baker--Akhiezer function $\psi_i$ as
\begin{align}
\label{eq:Kij}
K_{ij}=K(\xi_i,\xi_j)=\int_{-\infty}^x dx\psi_i\psi_j.
\end{align}
This means that
\begin{align}
\label{eq:dKij}
\partial_x K_{ij}=\psi_i\psi_j.
\end{align}
Let us introduce the notation
\begin{align}
\label{eq:bketkl}
\bket{k,l}
 &:=\int_{-\infty}^\infty d\xi_k\int_{-\infty}^\infty d\xi_{k+1}
 \cdots\int_{-\infty}^\infty d\xi_{l}
 e^{\sum_{j=k}^l\beta_j\xi_j}
 \psi_k K_{k,k+1}K_{k+1,k+2}\cdots K_{l-1,l}\psi_l
\end{align}
for $k\le l$. For $k=l$ it is understood that
\begin{align}
\bket{k,k}
 =\int_{-\infty}^\infty d\xi_k e^{\beta_k\xi_k}\psi_k^2
 =W_1(\beta_k).
\end{align}
With a slight abuse of notation let us also introduce
\begin{align}
\begin{aligned}
\bket{k,k-1}
 &:=\int_{-\infty}^\infty d\xi_1\cdots\int_{-\infty}^\infty d\xi_n
 \\
&\hspace{2em}
 \times
 e^{\sum_{j=1}^n\beta_j\xi_j}
 \psi_k K_{k,k+1}K_{k+1,k+2}
 \cdots K_{n-1,n}K_{n,1}K_{1,2}
 \cdots K_{k-2,k-1}\psi_{k-1}.
\end{aligned}
\end{align}
Using \eqref{eq:dKij}
we see that
\begin{align}
\tW_n(\beta_1,\beta_2,\ldots,\beta_n)
 =\sum_{k=1}^n \bket{k,k-1}.
\label{eq:tWinbket}
\end{align}
We find that $\bket{1,n}$ satisfies the simple differential equation
\begin{align}
\left(\partial_\tau-u\partial_x-\frac{1}{6}\partial_x^3\right)
\bket{1,n}
=-\sum_{1<k\le n}\bket{1,k-1}'\bket{k,n}'.
\label{eq:bketeq}
\end{align}
The proof of this equation is not difficult but rather lengthy,
so that we relegate it to Appendix~\ref{sec:proof}.
One can easily check that
\eqref{eq:bketeq} indeed reproduces the differential equations
\eqref{eq:W1KdV} and \eqref{eq:tW2KdV} for the $n=1,2$ cases
by observing that
\begin{align}
\bket{1,1}=W_1(\beta_1),\qquad
\bket{1,2}=\bket{2,1}=\frac{1}{2}\tW_2(\beta_1,\beta_2).
\end{align}

To compute the 't Hooft expansion of $\log \bket{1,n}$
we need to know the initial condition,
i.e.~the genus zero part.
This is derived as follows.
Recall that 
$K_{ij}$ is written as (see e.g.~\cite{Okuyama:2020ncd})
\begin{align}
\label{eq:Kijbis}
K_{ij}&=\frac{\psi_i'\psi_j-\psi_i\psi_j'}{\xi_i-\xi_j}.
\end{align}
Since the Baker--Akhiezer function admits the expansion
\eqref{eq:psiexp}, $K_{ij}$ is expanded as
\begin{align}
\label{eq:Kijexp}
K_{ij}=e^{\frac{A_0(\xi_i)+A_0(\xi_j)}{\gs}+{\cal O}(\gs^0)}.
\end{align}
Plugging \eqref{eq:psiexp} and \eqref{eq:Kijexp} into \eqref{eq:bketkl}
one obtains
\begin{align}
\label{eq:bketinit}
\bket{k,l}=e^{\frac{1}{\gs}\sum_{j=k}^l G_0(s_j)+{\cal O}(\gs^0)}.
\end{align}

By solving the differential equation \eqref{eq:bketeq}
with the initial condition \eqref{eq:bketinit}
one can determine
the 't Hooft expansion of $\bket{1,n}$
purely algebraically, as in the case of $\tW_2$.
Then, using the relation \eqref{eq:tWinbket},
one immediately obtains the 't Hooft expansion of $\tW_n$
and $\tZ_n$.

\section{'t Hooft expansion of the three-boundary correlator}\label{sec:3pt}

\subsection{General results}

In this section let us compute the 't Hooft expansion of
the three-point function using the formalism developed
in the last section.
We consider the 't Hooft expansion
\begin{align}
\cK^{(3)}
 =\log\tZ_3
 =\sum_{\dg=0}^\infty \gs^{\dg-1} \cK^{(3)}_\dg.
\label{eq:K3exp}
\end{align}
As we mentioned, one can algebraically
compute the above genus expansion. The results are
\begin{align}
\begin{aligned}
\cK^{(3)}_0&=\cF_0(s_1)+\cF_0(s_2)+\cF_0(s_3),\\
\cK^{(3)}_1
 &=\frac{1}{2}
   \log\left[\xi_{1*}^{(1)}\xi_{2*}^{(1)}\xi_{3*}^{(1)}\right]
  -\log\left[z_{1*}z_{2*}z_{3*}
             (z_{1*}+z_{2*})(z_{2*}+z_{3*})(z_{3*}+z_{1*})\right]
  +\ln 2,\\
\cK^{(3)}_2
 &=\biggl(
 -\frac{5}{12tz_{1*}^3}
 -\frac{1}{2tz_{1*}^2}\left[\frac{1}{z_{2*}}+\frac{1}{z_{3*}}\right]
 -\frac{I_2}{12t^2z_{1*}}
 +\frac{\xi_{1*}^{(3)}}{16\bigl(\xi_{1*}^{(1)}\bigr)^2}
 -\frac{\bigl(\xi_{1*}^{(2)}\bigr)^2}
       {12\bigl(\xi_{1*}^{(1)}\bigr)^3}\\
&\hspace{1.8em}
 -\frac{\xi_{1*}^{(2)}}{4\xi_{1*}^{(1)}}
  \left[\frac{1}{z_{1*}^2}+\frac{1}{z_{1*}(z_{1*}+z_{2*})}
                          +\frac{1}{z_{1*}(z_{1*}+z_{3*})}\right]\\
&\hspace{1.8em}
 +\frac{\xi_{1*}^{(1)}}{4}
  \left[\frac{3}{z_{1*}^4}+\frac{3}{z_{1*}^3(z_{1*}+z_{2*})}
                          +\frac{3}{z_{1*}^3(z_{1*}+z_{3*})}\right.\\
&\hspace{5.5em}\left.
       +\frac{2}{z_{1*}^2(z_{1*}+z_{2*})^2}
       +\frac{2}{z_{1*}^2(z_{1*}+z_{3*})^2}
       +\frac{2}{z_{1*}^2(z_{1*}+z_{2*})(z_{1*}+z_{3*})}
  \right]\biggr)\\
 &\hspace{1em}+\mbox{cyclic perm.}
\end{aligned}
\end{align}
We checked that the above $\cK^{(3)}_1$ is in agreement
(up to a constant) with the saddle-point result
calculated in \cite{Okuyama:2020ncd}.
We also verified that the above $\cK^{(3)}_2$
restricted to the JT gravity case $t_k=\gamma_k\ (k\ge 0)$
is in perfect agreement with $2^{-1/2}\cK^{(3)}_2\Big|_{y=0,t=1}$
of \cite{Okuyama:2020ncd}.
Note that in the case of $t_k=\gamma_k\ (k\ge 0)$ we have
\begin{align}
\begin{aligned}
t&=1,\qquad I_2=1,\\
\xi_{i*}^{(1)}
 &=\frac{z_{i*}}{\cos\left(\sqrt{2}z_{i*}\right)},\qquad
\xi_{i*}^{(2)}
 =\frac{\sqrt{2}z_{i*}\sin\left(\sqrt{2}z_{i*}\right)
        +\cos\left(\sqrt{2}z_{i*}\right)}
       {\cos^3\left(\sqrt{2}z_{i*}\right)},\\[1ex]
\xi_{i*}^{(3)}
 &=\frac{-4z_{i*}\cos^2\left(\sqrt{2}z_{i*}\right)
         +3\sqrt{2}\sin\left(\sqrt{2}z_{i*}\right)
                   \cos\left(\sqrt{2}z_{i*}\right)
         +6z_{i*}}
        {\cos^5\left(\sqrt{2}z_{i*}\right)}.
\end{aligned}
\end{align}
Note also that
the $z_{i*}$ terms here are related to those in \cite{Okuyama:2020ncd}
by $z_{i*}^{\rm here}=\sqrt{2}z_{i*}^{\rm there}$.

\subsection{Airy case}

In this subsection we consider the Airy case 
corresponding to a particular subspace of couplings $\{t_n\}$
\begin{equation}
\begin{aligned}
 t_0,t_1\ne0,\qquad t_k=0~~(k\geq2).
\end{aligned} 
\label{eq:airy-sub}
\end{equation}
In this case the Itzykson--Zuber variables $I_n$ in \eqref{eq:defIn} become
\begin{align}
\begin{aligned}
I_0&=t_0+u_0t_1,\qquad I_1=t_1,\qquad I_{k\ge 2}=0,
\end{aligned}
\label{eq:In-airy}
\end{align}
and from the genus-zero string equation $u_0=I_0$ we find
\begin{equation}
\begin{aligned}
 u_0&=\frac{t_0}{t},\qquad t=1-t_1.
\end{aligned} 
\label{eq:ut-airy}
\end{equation}
From \eqref{eq:In-airy} and \eqref{eq:ut-airy}, one
can see that various quantities introduced in section \ref{sec:1pt} take very simple 
form
\begin{align}
\begin{aligned}
\Fo_0&=u_0 s+\frac{s^3}{6t^2},\qquad
z_{*}=\frac{s}{t},\qquad
\xi_{*}=u_0+\frac{s^2}{2t^2},\\
\xi_{*}^{(1)}&=\frac{s}{t^2}=\frac{z_{*}}{t},\qquad
\xi_{*}^{(2)}=\frac{1}{t^2},\qquad
\xi_{*}^{(k\ge 3)}=0.
\end{aligned}
\end{align}
Using these we obtain the 't Hooft expansion \eqref{eq:K3exp} of $\cK^{(3)}$
in the Airy case. For instance, $\cK^{(3)}_2$ and $\cK^{(3)}_3$ are given by
\begin{align}
\begin{aligned}
\cK^{(3)}_2
 &=-\frac{z_{1*}^3+z_{2*}^3+z_{3*}^3+3z_{1*}z_{2*}z_{3*}}
 {2tz_{1*}z_{2*}z_{3*}(z_{1*}+z_{2*})(z_{2*}+z_{3*})(z_{3*}+z_{1*})},
\\[1ex]
\cK^{(3)}_3
 &=\frac{1}{8t^2\left[z_{1*}z_{2*}z_{3*}
   (z_{1*}+z_{2*})(z_{2*}+z_{3*})(z_{3*}+z_{1*})\right]^2}\\[1ex]
 &\hspace{1em}
 \times[
  5\left(z_{1*}^6+z_{2*}^6+z_{3*}^6\right)
 +6\left((z_{2*}+z_{3*})z_{1*}^5
        +(z_{3*}+z_{1*})z_{2*}^5
        +(z_{1*}+z_{2*})z_{3*}^5\right)\\
 &\hspace{2.5em}
 +14\left(z_{2*}z_{3*}z_{1*}^4
         +z_{3*}z_{1*}z_{2*}^4
         +z_{1*}z_{2*}z_{3*}^4\right)
 -2\left(z_{1*}^3z_{2*}^3
        +z_{2*}^3z_{3*}^3
        +z_{3*}^3z_{1*}^3\right)\\
 &\hspace{2.5em}
 +21z_{1*}^2 z_{2*}^2 z_{3*}^2
 ].
\end{aligned}
\label{eq:K3-airy}
\end{align}

In the Airy case, it is known that multi-point functions
$Z_n(\bt_1,\ldots,\bt_n)$ can be written in the integral representation
\cite{okounkov2002generating}.
In particular, we can write down
$Z_n$ for $n=1,2$ in a closed form \cite{okounkov2002generating}\footnote{Strictly speaking,
the closed form of $Z_{1,2}$ is obtained in \cite{okounkov2002generating}
when $t_n=0~(n\geq0)$, but it is straightforward to generalize
the result in \cite{okounkov2002generating} to the case 
of our interest \eqref{eq:airy-sub}.}
\begin{equation}
\begin{aligned}
 Z_1(\bt)&=\frac{1-t_1}{\gs\rt{2\pi\bt^3}}\exp\left(\frac{\bt t_0}{1-t_1}+\frac{\gs^2\bt^3}{24(1-t_1)^2}\right),\\
Z_2(\bt_1,\bt_2)&=Z_1(\bt_1+\bt_2)\text{Erf}\left(\frac{\gs\rt{\bt_2\bt_3(\bt_2+\bt_3)}}{2\rt{2}(1-t_1)}\right),
\end{aligned} 
\label{eq:z12-airy}
\end{equation}
where $\text{Erf}(z)$ denotes the error function
\begin{equation}
\begin{aligned}
 \text{Erf}(z)=\frac{2}{\rt{\pi}}\int_0^zdx e^{-x^2}.
\end{aligned} 
\end{equation}
For $n=3$, the closed form expression of
$Z_3(\bt_1,\bt_2,\bt_3)$ with $\bt_1=\bt_2=\bt_3$
is obtained in \cite{Beccaria:2020ykg} in terms of the
Owen's $T$-function 
\begin{equation}
\begin{aligned}
 T(z,a)=\frac{1}{2\pi}\int_0^a dx\frac{e^{-\hf z^2(1+x^2)}}{1+x^2}.
\end{aligned} 
\end{equation}
We can generalize the result of \cite{Beccaria:2020ykg}
for $Z_3(\bt_1,\bt_2,\bt_3)$ with arbitrary $\bt_1,\bt_2,\bt_3$.
As discussed in \cite{Okuyama:2021pkf}, one can determine 
$Z_3$ by solving the KdV equation for $W_3$ \eqref{eq:WnKdV} on the subspace
\eqref{eq:airy-sub}
using $Z_{1,2}$ in \eqref{eq:z12-airy} as inputs.
In this way we find
\begin{equation}
\begin{aligned}
  \frac{Z_3(\bt_1,\bt_2,\bt_3)}{Z_1(\bt_1+\bt_2+\bt_3)}
=1&
-4T\left(\frac{\gs\rt{\bt_1(\bt_2+\bt_3)(\bt_1+\bt_2+\bt_3)}}{2(1-t_1)},\rt{\frac{\bt_2\bt_3}{\bt_1(\bt_1+\bt_2+\bt_3)}}\right)\\
&-4T\left(\frac{\gs\rt{\bt_2(\bt_3+\bt_1)(\bt_1+\bt_2+\bt_3)}}{2(1-t_1)},\rt{\frac{\bt_3\bt_1}{\bt_2(\bt_1+\bt_2+\bt_3)}}\right)\\
&-4T\left(\frac{\gs\rt{\bt_3(\bt_1+\bt_2)(\bt_1+\bt_2+\bt_3)}}{2(1-t_1)},\rt{\frac{\bt_1\bt_2}{\bt_3(\bt_1+\bt_2+\bt_3)}}\right).
\end{aligned} 
\label{eq:airy-z3}
\end{equation}
One can easily see that this reduces to the result of \cite{Beccaria:2020ykg}
when $\bt_1=\bt_2=\bt_3$.

From this exact result \eqref{eq:airy-z3} of $Z_3$,
one can compute the 't Hooft expansion \eqref{eq:K3exp}
of $\cK^{(3)}=\log\til{Z}_3$ in the Airy case.
Using the relation between $Z_3$ and $\til{Z}_3$
in \eqref{eq:Zn-tZn} and the following property of the Owen's $T$-function
\begin{equation}
\begin{aligned}
 T(z,\infty)=\qu-\qu\text{Erf}\left(\frac{z}{\rt{2}}\right),
\end{aligned} 
\end{equation}
we find
\begin{equation}
\begin{aligned}
 \frac{\til{Z}_3(\bt_1,\bt_2,\bt_3)
}{Z_1(\bt_1+\bt_2+\bt_3)}
=&2\til{T}\left(\frac{\gs\rt{\bt_1(\bt_2+\bt_3)(\bt_1+\bt_2+\bt_3)}}{2(1-t_1)},\rt{\frac{\bt_2\bt_3}{\bt_1(\bt_1+\bt_2+\bt_3)}}\right)\\
+&2\til{T}\left(\frac{\gs\rt{\bt_2(\bt_3+\bt_1)(\bt_1+\bt_2+\bt_3)}}{2(1-t_1)},\rt{\frac{\bt_3\bt_1}{\bt_2(\bt_1+\bt_2+\bt_3)}}\right)\\
+&2\til{T}\left(\frac{\gs\rt{\bt_3(\bt_1+\bt_2)(\bt_1+\bt_2+\bt_3)}}{2(1-t_1)},\rt{\frac{\bt_1\bt_2}{\bt_3(\bt_1+\bt_2+\bt_3)}}\right),
\end{aligned} 
\label{eq:tz3-airy}
\end{equation}
where we defined
\begin{equation}
\begin{aligned}
 \til{T}(z,a)=T(z,\infty)-T(z,a)=\frac{1}{2\pi}\int_a^{\infty}dx
\frac{e^{-\hf z^2(1+x^2)}}{1+x^2}.
\end{aligned} 
\end{equation}
In the large $z$ regime with finite $a$, this is expanded as
\begin{equation}
\begin{aligned}
 \til{T}(z,a)&= \frac{e^{-\hf(1+a^2)z^2}}{2\pi}\sum_{k=0}^\infty z^{-2(k+1)} 
\Bigl(\frac{1}{a}\del_a\Bigr)^k
\frac{1}{a(1+a^2)}\\
&= \frac{e^{-\hf(1+a^2)z^2}}{2\pi }\left[\frac{1}{a(1+a^2)z^2}-\frac{1+3a^2}{a^3(1+a^2)^2z^4}
+\frac{3+10a^2+15a^4}{a^5(1+a^2)^3z^6} +\cdots\right].
\end{aligned} 
\end{equation}
Using this expansion, we checked that the 't Hooft expansion of 
the exact result of $\til{Z}_3$ in \eqref{eq:tz3-airy}
reproduces $\cK_2^{(3)}$ and $\cK_3^{(3)}$ in \eqref{eq:K3-airy}.
This serves as a nontrivial consistency check of our formalism.

\section{Conclusions and outlook}\label{sec:conclusion}
In this paper we developed a formalism to compute 
the 't Hooft expansion of the multi-boundary correlators in topological gravity
with general couplings $t_k$.
The 't Hooft expansion of the $n$-point function can be obtained from the relation
\eqref{eq:bketeq}, which is equivalent to \eqref{eq:WnKdV} for $n=1,2$.
For the one-point function, we developed an algorithm for
the computation of 't Hooft expansion in section \ref{sec:1pt}, 
which is almost parallel to the computation of open free energy 
studied in \cite{Okuyama:2020vrh}.
We find that the 't Hooft expansion of an $n$-point function $(n\geq2)$
is determined algebraically from the lower point functions
by using \eqref{eq:bketeq}.
Our computation reproduces the JT gravity case studied in
\cite{Okuyama:2019xbv,Okuyama:2020ncd} and the 't Hooft expansion of the exact result
of correlators in the Airy case, as it should.

There are several interesting open questions.
It is suggested in \cite{Liu:2018hlr} that we can define the multi-point analogue
of the spectral form factor in random matrix models and
it exhibits a similar behavior as the ramp and plateau.
Using our formalism, it would be possible to
study the ramp--plateau transition regime of the 
multi-point version of the spectral form factor.
We leave this as an interesting future problem.

Recently, the quenched free energy $\bra \log Z(\bt)\ket$
of JT gravity is studied by the replica method 
\cite{Engelhardt:2020qpv,Johnson:2020mwi,Alishahiha:2020jko}.
It is shown in \cite{Okuyama:2021pkf} that the quenched free energy
is written as a certain integral transformation of the generating function
of multi-boundary correlators. The low-temperature behavior of 
quenched free energy in JT gravity is quite interesting since
it is suggested in \cite{Engelhardt:2020qpv} that JT gravity exhibits 
a spin glass phase at low temperature.
In \cite{Engelhardt:2020qpv,Okuyama:2021pkf}
the quenched free energy is analyzed in the Airy regime
$\bt\sim\gs^{-2/3}$. It would be interesting to study
the quenched free energy of JT gravity
in the 't Hooft regime
$\bt\sim\gs^{-1}$ using our formalism.

\acknowledgments
This work was supported in part by JSPS KAKENHI Grant
Nos.~19K03845 and 19K03856,
and JSPS Japan-Russia Research Cooperative Program.

\appendix

\section{Proof of the master differential equation}\label{sec:proof}

In this section we prove \eqref{eq:bketeq} for general $n$.
Let us introduce the notation
\begin{align}
\bket{k^{(p)},l^{(q)}}
 &:=\int_{-\infty}^\infty d\xi_k\int_{-\infty}^\infty d\xi_{k+1}
 \cdots\int_{-\infty}^\infty d\xi_{l}
 e^{\sum_{j=k}^l\beta_j\xi_j}
 \partial_x^p\psi_k K_{k,k+1}K_{k+1,k+2}\cdots K_{l-1,l}
 \partial_x^q\psi_l.
\end{align}
Derivatives of $\bket{1,n}$ are
calculated as
\begin{align}
\begin{aligned}
\bket{1,n}'
 &=\sum_{1<k\le n}\bket{1,k-1}\bket{k,n}
 +\left(\bket{1',n}+\bket{1,n'}\right),\\
\bket{1,n}''
 &=2\sum_{1<k<l\le n}\bket{1,k-1}\bket{k,l-1}\bket{l,n}\\
&\hspace{1em}
 +\sum_{1<k\le n}\left(2\bket{1',k-1}\bket{k,n}
 +\bket{1,k-1'}\bket{k,n}\right.\\[-1ex]
&\hspace{4.5em}
 +\left.\bket{1,k-1}\bket{k',n}
 +2\bket{1,k-1}\bket{k,n'}\right)\\[1ex]
&\hspace{1em}
 +\left(\bket{1'',n}+\bket{1,n''}\right)
 +2\bket{1',n'},\\[1ex]
\bket{1,n}'''
 &=6\sum_{1<k<l<m\le n}\bket{1,k-1}\bket{k,l-1}\bket{l,m-1}\bket{m,n}\\
&\hspace{1em}
 +\sum_{1<k<l\le n}\left(6\bket{1',k-1}\bket{k,l-1}\bket{l,n}
 +3\bket{1,k-1'}\bket{k,l-1}\bket{l,n}\right.\\[-1ex]
&\hspace{6em}
 +3\left.\bket{1,k-1}\bket{k',l-1}\bket{l,n}
 +3\bket{1,k-1}\bket{k,l-1'}\bket{l,n}\right.\\[1.5ex]
&\hspace{6em}
 +3\left.\bket{1,k-1}\bket{k,l-1}\bket{l',n}
 +6\bket{1,k-1}\bket{k,l-1}\bket{l,n'}\right)\\[1ex]
&\hspace{1em}
 +\sum_{1<k\le n}\left(3\bket{1'',k-1}\bket{k,n}
 +\bket{1,k-1''}\bket{k,n}\right.\\[-1ex]
&\hspace{4.5em}
 +\left.\bket{1,k-1}\bket{k'',n}
 +3\bket{1,k-1}\bket{k,n''}\right.\\[1ex]
&\hspace{4.5em}
 +\left. 6\bket{1',k-1}\bket{k,n'}
 +2\bket{1,k-1'}\bket{k',n}\right.\\[1ex]
&\hspace{4.5em}
 +\left. 3\bket{1',k-1}\bket{k',n}
 +3\bket{1,k-1'}\bket{k,n'}\right.\\[1ex]
&\hspace{4.5em}
 +\left. 3\bket{1',k-1'}\bket{k,n}
 +3\bket{1,k-1}\bket{k',n'}\right)\\[1ex]
&\hspace{1em}
 +\left(\bket{1''',n}+\bket{1,n'''}\right)
 +3\left(\bket{1'',n'}+\bket{1',n''}\right).
\end{aligned}
\label{eq:bket1ndiff}
\end{align}

Next, from \eqref{eq:auxlin} we see that
\begin{align}
\label{eq:cdpsi}
\begin{aligned}
\left(\partial_\tau-u\partial_x\right)\psi_i
 &=\frac{2}{3}\psi_i'''+\frac{1}{2}u'\psi_i\\
 &=\frac{1}{6}\psi_i'''
  +\frac{1}{2}\partial_x(\partial_x^2+u)\psi_i-\frac{1}{2}u\psi_i'\\
 &=\frac{1}{6}\psi_i'''+\frac{1}{2}(\xi_i-u)\psi_i'.
\end{aligned}
\end{align}
It also follows that
\begin{align}
\begin{aligned}
\dot{K}_{ij}
 &=\int_{-\infty}^x dx (\dot{\psi}_i\psi_j+\psi_i\dot{\psi}_j)\\
 &=\int_{-\infty}^x dx \left[
  \frac{2}{3}(\psi_i'''\psi_j+\psi_i\psi_j''')
  +u(\psi_i'\psi_j+\psi_i\psi_j')+u'\psi_i\psi_j\right]\\
 &=\frac{2}{3}(\psi_i''\psi_j+\psi_i\psi_j''-\psi_i'\psi_j')
  +u\psi_i\psi_j
\end{aligned}
\end{align}
and thus
\begin{align}
\label{eq:cdK}
\begin{aligned}
\left(\partial_\tau-u\partial_x\right)K_{ij}
 &=\frac{2}{3}(\psi_i''\psi_j+\psi_i\psi_j''-\psi_i'\psi_j').
\end{aligned}
\end{align}
Using \eqref{eq:cdpsi} and \eqref{eq:cdK}, we obtain
\begin{align}
\label{eq:cdbket1}
\begin{aligned}
\left(\partial_\tau-u\partial_x\right)\bket{1,n}
 &=
 \frac{2}{3}\sum_{1<k\le n}\left(
  \bket{1,k-1''}\bket{k,n}
 +\bket{1,k-1}\bket{k'',n}
 -\bket{1,k-1'}\bket{k',n}\right)\\
&\hspace{1em}
 +\frac{1}{6}\left(\bket{1''',n}+\bket{1,n'''}\right)
 +\frac{1}{2}\left((\xi_1-u)\bket{1',n}
  +(\xi_n-u)\bket{1,n'}\right).
\end{aligned}
\end{align}
Note that \eqref{eq:Kijbis} is rewritten as
\begin{align}
(\xi_i-\xi_j)K_{ij}=\psi_i'\psi_j-\psi_i\psi_j'
\end{align}
and thus
\begin{align}
(\xi_i-u)K_{ij}=(\xi_j-u)K_{ij}+\psi_i'\psi_j-\psi_i\psi_j'.
\label{eq:Kpsirel}
\end{align}
Using this we obtain
\begin{align}
\begin{aligned}
(\xi_1-u)\bket{1',n}
 &=(\xi_n-u)\bket{1',n}
  +\sum_{1<k\le n}
   \left(\bket{1',k-1'}\bket{k,n}-\bket{1',k-1}\bket{k',n}\right),\\
 &=\bket{1',n''}
  +\sum_{1<k\le n}
   \left(\bket{1',k-1'}\bket{k,n}-\bket{1',k-1}\bket{k',n}\right),\\
(\xi_n-u)\bket{1,n'}
 &=\bket{1'',n'}
  +\sum_{1<k\le n}
   \left(\bket{1,k-1}\bket{k',n'}-\bket{1,k-1'}\bket{k,n'}\right).
\end{aligned}
\end{align}
\eqref{eq:cdbket1} is then written as
\begin{align}
\label{eq:cdbket2}
\begin{aligned}
\left(\partial_\tau-u\partial_x\right)\bket{1,n}
 &=
 \frac{2}{3}\sum_{1<k\le n}\left(
  \bket{1,k-1''}\bket{k,n}
 +\bket{1,k-1}\bket{k'',n}
 -\bket{1,k-1'}\bket{k',n}\right)\\
&\hspace{1em}
 +\frac{1}{6}\left(\bket{1''',n}+\bket{1,n'''}\right)
 +\frac{1}{2}\left(\bket{1',n''}+\bket{1'',n'}\right)\\[1ex]
&\hspace{1em}
 +\frac{1}{2}\sum_{1<k\le n}
   \left(\bket{1',k-1'}\bket{k,n}-\bket{1',k-1}\bket{k',n}\right.\\
&\hspace{5em}
  +\left.\bket{1,k-1}\bket{k',n'}-\bket{1,k-1'}\bket{k,n'}\right).
\end{aligned}
\end{align}

Combining \eqref{eq:bket1ndiff} and \eqref{eq:cdbket2}, we obtain
\begin{align}
\begin{aligned}
\lefteqn{
\left(\partial_\tau-u\partial_x-\frac{1}{6}\partial_x^3\right)
\bket{1,n}}\\
 &=\frac{1}{2}\sum_{1<k\le n}\left(
  \bket{1,k-1''}\bket{k,n}
 +\bket{1,k-1}\bket{k'',n}\right.\\[-1ex]
&\hspace{5em}\left.
 -\bket{1'',k-1}\bket{k,n}
 -\bket{1,k-1}\bket{k,n''}\right)\\[1ex]
&\hspace{1em}
 -\sum_{1<k\le n}
   \left(\bket{1',k-1}+\bket{1,k-1'}\right)
   \left(\bket{k',n}+\bket{k,n'}\right)\\
&\hspace{1em}
 +\sum_{1<k<l\le n}\left(-\bket{1',k-1}\bket{k,l-1}\bket{l,n}
 -\frac{1}{2}\bket{1,k-1'}\bket{k,l-1}\bket{l,n}\right.\\[-1ex]
&\hspace{6em}\left.
 -\frac{1}{2}\bket{1,k-1}\bket{k',l-1}\bket{l,n}
 -\frac{1}{2}\bket{1,k-1}\bket{k,l-1'}\bket{l,n}\right.\\[1.5ex]
&\hspace{6em}\left.
 -\frac{1}{2}\bket{1,k-1}\bket{k,l-1}\bket{l',n}
 -\bket{1,k-1}\bket{k,l-1}\bket{l,n'}\right)\\[1ex]
&\hspace{1em}
 -\sum_{1<k<l<m\le n}\bket{1,k-1}\bket{k,l-1}\bket{l,m-1}\bket{m,n}.
\end{aligned}
\end{align}
The first term of the r.h.s.~can be rewritten by using the relations
\begin{align}
\begin{aligned}
\bket{1,k-1''}-\bket{1'',k-1}
&=(\xi_{k-1}-u)\bket{1,k-1}-(\xi_1-u)\bket{1,k-1}\\[1ex]
&=\sum_{1<j<k}(\xi_j-\xi_{j-1})\bket{1,k-1}\\
&=\sum_{1<j<k}
  \left(-\bket{1,j-1'}\bket{j,k-1}+\bket{1,j-1}\bket{j',k-1}\right),\\
\bket{k'',n}-\bket{k,n''}
&=(\xi_k-u)\bket{1,k-1}-(\xi_n-u)\bket{k,n}\\[1ex]
&=\sum_{k<l\le n}(\xi_{l-1}-\xi_l)\bket{k,n}\\
&=\sum_{k<l\le n}
  \left(\bket{k,l-1'}\bket{l,n}-\bket{k,l-1}\bket{l',n}\right).
\end{aligned}
\end{align}
Hence we obtain
\begin{align}
\begin{aligned}
\lefteqn{
\left(\partial_\tau-u\partial_x-\frac{1}{6}\partial_x^3\right)
\bket{1,n}}\\
&=-\sum_{1<k\le n}
   \left(\bket{1',k-1}+\bket{1,k-1'}\right)
   \left(\bket{k',n}+\bket{k,n'}\right)\\
&\hspace{1em}
 +\sum_{1<k<l\le n}\left(-\bket{1',k-1}\bket{k,l-1}\bket{l,n}
 -\bket{1,k-1'}\bket{k,l-1}\bket{l,n}\right.\\[-1ex]
&\hspace{6em}\left.
 -\bket{1,k-1}\bket{k,l-1}\bket{l',n}
 -\bket{1,k-1}\bket{k,l-1}\bket{l,n'}\right)\\[1ex]
&\hspace{1em}
 -\sum_{1<k<l<m\le n}\bket{1,k-1}\bket{k,l-1}\bket{l,m-1}\bket{m,n}\\
&=-\sum_{1<k\le n}
   \Bigl(\bket{1',k-1}+\bket{1,k-1'}
    +\sum_{1<j<k}\bket{1,j-1}\bket{j,k-1}\Bigr)\\
&\hspace{3.5em}\times
   \Bigl(\bket{k',n}+\bket{k,n'}
    +\sum_{k<l\le n}\bket{k,l-1}\bket{l,n}\Bigr)\\
&=-\sum_{1<k\le n}\bket{1,k-1}'\bket{k,n}'.
\end{aligned}
\end{align}
%

\bibliography{paper}
\bibliographystyle{utphys}

\end{document}